\documentclass{ws-ijmpcs}

\begin{document}

\markboth{S.N. Nedelko \& V.E. Voronin}
{Radial Meson Excitations and Abelian Self-Dual Gluon Fields}

%
\catchline{}{}{}{}{}
%

\title{ Radial Meson Excitations and Abelian Self-Dual Gluon Fields}
\author{Sergei N. Nedelko}

\address{ Bogoliubov Laboratory of Theoretical Physics, JINR,
141980 Dubna, Russia \\
nedelko@theor.jinr.ru}

\author{Vladimir E. Voronin}

\address{Bogoliubov Laboratory of Theoretical Physics, JINR,
141980 Dubna, Russia\\
nedelko@theor.jinr.ru}

\maketitle

\begin{history}
\received{Day Month Year}
\revised{Day Month Year}
\published{Day Month Year}
\end{history}

\begin{abstract}
We briefly review  motivation and  results of the approach to QCD vacuum as a medium describable in terms of statistical ensemble of almost everywhere  constant Abelian (anti-)self-dual  gluon fields.  An overview of the hadronization formalism based on this ensemble  is presented. 
 New results for radial excitations of light, heavy-light mesons and heavy quarkonia are presented. A possible interrelation between  the present approach and  holographic QCD with harmonic confinement  is outlined.

\end{abstract}

\ccode{PACS numbers: 12.38.Aw, 12.38.Lg, 12.38.Mh, 11.15.Tk}

\section{Introduction}
 In recent years,  the approaches to confinement based on the ideas of the soft wall ADS/QCD model and light front holography demonstrated an impressive phenomenological success~\cite{qds/qcd,brodsky}. The crucial  features of these approaches are the particular dilaton profile $\varphi(z)=\kappa^2 z^2$ and the harmonic oscillator form of the confining potential as the function of  fifth coordinate $z$  .  In this respect  holographic QCD can be seen as relative to the formalism developed by Leutwyler and Stern  in a series of papers devoted to the  covariant description of bilocal meson-like fields $\Phi(x,z)$ combined with the idea of harmonic confinement~\cite{leutwyler-stern}. 

In this paper we briefly review the approach to QCD bosonization based on the   representation of  QCD vacuum  in terms of the statistical ensemble of  almost everywhere homogeneous Abelian (anti-)self-dual gluon fields,  calculate the spectrum of radial  meson excitations and their decay constants.  We outline  the  possible relation of the soft wall AdS/QCD and light-front holography ideas with this, at first sight,  completely different approach.
 The  character of meson wave functions   in our approach~\cite{EN1} is predetermined by the form of the gluon propagator in the  background  of vacuum gluon field.   These wave functions  are almost identical to the wave functions of the soft wall AdS/QCD with  quadratic dilaton profile. In both cases we are dealing with the generalized Laguerre polynomials as the functions of $z$. In the hadronization approach of Refs.~\refcite{EN1} and \refcite{EN}  the fifth coordinate $z$ appears as the relative distance between quark and antiquark in the center of mass coordinate system, while the  center of mass coordinate $x$ represents the space-time point where the extended meson field is localized.

The physical origin for the particular form of dilaton profile in ADC/QCD and light front holography as well as the harmonic potential in the Stern-Leutwyler consideration, and, hence, for the Laguerre polynomial form of the meson wave functions, could  not be identified within these approaches themselves. On the contrary, the approach based on bosonization in the background gluon fields links the required by the meson phenomenology form of the meson wave functions to a particular picture of QCD vacuum. 

The paper is organized as follows. Section \ref{section_domain_wall_network_as_vacuum} is devoted to motivation of the approach.  Effective meson action  is considered in section \ref{section_domain_model}.  Results for the  masses, transition and decay constants of various mesons are presented in  section \ref{section_masses_of_mesons}. In the  last section  we outline possible relation between hadronization approach and the formalism of the soft wall AdS/QCD model.

\section{Domain wall network as QCD vacuum \label{section_domain_wall_network_as_vacuum}}

 Functional integral approach to quantization of QCD requires certain  conditions defining  the functional space of fields to be integrated over (see, e.g., Ref.\refcite{faddeev}). In the very basic 
  representation of the Euclidean  functional integral for QCD, 
\begin{equation}
\label{functional_integral}
Z=N\int\limits_{{\cal F}_B} DA \int\limits_{\Psi} D\psi D\bar\psi\exp\{-S[A,\psi,\bar{\psi}]\},
\end{equation}
integration spaces $\mathcal{F}_B$ for gluon and  $\Psi$ for quark fields  have to be specified.
Bearing in mind a nontrivial QCD vacuum structure encoded in various condensates, one have to define  $\mathcal{F}_B$ allowing  gluon fields with nonzero classical action density,
\begin{equation}
\label{condensate}
{\cal F}_B=\left\{A: \lim_{V\to \infty}\frac{1}{V}\int_V d^4xg^2F^a_{\mu\nu} (x)F^a_{\mu\nu}(x) =B^2\right\}.
\end{equation}
The  value of $B$  is allowed to be nonzero. Separation of modes  $B_\mu^a$ responsible for nonzero  condensates  from the small perturbations $Q_\mu^a$ must be supplemented with gauge fixing.  Background gauge fixing condition $D(B)Q=0$ is the most natural choice. To perform separation, one inserts identity
\begin{equation}
1=\int\limits_{{\cal B}}DB \Phi[A,B]\int\limits_{{\cal Q}} DQ\int\limits_{\Omega}D\omega \delta[A^\omega-Q^\omega-B^\omega]
 \delta[D(B^\omega)Q^\omega]
\end{equation}
in the functional integral and arrives at 
\begin{eqnarray*}
\label{effpot}
Z &=&N'\int\limits_{{\cal B}}DB \int\limits_{\Psi} D\psi D\bar\psi\int\limits_{{\cal Q}} DQ \det[\mathcal{D}(B)\mathcal{D}(B+Q)]
\delta[\mathcal{D}(B)Q]e^{-S_{\rm QCD}[B+Q,\psi, \bar\psi]}\\
&=&\int\limits_\mathcal{B}DB \exp\{-S_\mathrm{eff}[B]\}.
\end{eqnarray*}
Thus defined quantum effective action $S_\text{eff}[B]$ has a physical meaning of the free energy of the quantum field system in the presence of the background  gluon field $B_\mu ^a$.  In the limit $V\to \infty$  global minima of  $S_\text{eff}[B]$ determine the class of  gauge field configurations  representing  the equilibrium state (vacuum ) of the system. 

Quite reliable argumentation in favor of (almost everywhere) homogeneous Abelian (anti-)self-dual fields as dominating vacuum configurations was put forward by many authors.  Pagels and Tomboulis showed\cite{Pagels}  that  these background fields represented a medium infinitely stiff to small color charged field fluctuations. This feature was interpreted as suggestive for confinement of color.
A comprehensive analysis of homogeneous fields was done by Leutwyler \cite{Leutwyler2}. Abelian (anti-)self-dual covariantly constant fields were shown to be the only constant fields free from tachyonic modes. Furthermore, propagators of color charged fields in constant background Abelian  (anti-)self-dual field are entire analytical functions of  momentum, which has been treated as confinement of color \cite{Leutwyler1}.
Calculation of QCD quantum effective action  in functional renormalization group approach \cite{Pawlowski} supported the results of one-loop calculations \cite{Pagels,Minkowski,Leutwyler2} and indicated the existence of a minimum of the effective potential for nonzero value of Abelian (anti-)self-dual homogeneous gluon field .

 Ginzburg-Landau (GL) approach to the quantum effective action indicated a possibility of the domain wall network formation in  QCD vacuum resulting in the  the dominating vacuum gluon configuration seen as an ensemble of densely  packed lumps of Abelian (anti-)self-dual field \cite{NG2011}. This conclusion is based on the observation that assumption about the nonzero scalar gluon condensate leads to existence  of twelve discrete degenerate global minima of the effective action. 
\begin{equation*}
A^{k,\pm}_{\mu}  = -\frac{1}{2}n_k F_{\mu\nu}x_\nu, \, \tilde F_{\mu\nu}=\pm F_{\mu\nu},
\end{equation*}
where $"\pm"$ denotes (anti-)self-duality and  matrix $\breve{n}_k$ belongs to Cartan subalgebra of $su(3)$
\begin{gather*}
 n_k = T^3\ \cos\left(\xi_k\right) + T^8\ \sin\left(\xi_k\right),
\ \
\xi_k=\frac{2k+1}{6}\pi, \, k=0,1,\dots,5.
\end{gather*}
The minima are connected by the parity and Weyl group reflections.  
Their existence indicates that the system is prone to the domain wall formation.  To demonstrate the simplest example of domain wall interpolating between the self-dual and anti-self-dual Abelian configurations, one allows  the angle  $\omega$ between chromomagnetic and chromoelectric fields to vary and restricts other degrees of freedom of gluon field to their vacuum values.
In this case GL Lagrangian  leads  to the  sine-Gordon equation for  $\omega$ with the standard
 kink solution (for details see Ref.~\refcite{NG2011})
\begin{equation}
 \omega(x_\nu) = 2\ \arctan \left(\exp(\mu x_\nu)\right).
\end{equation}

\begin{figure}[h!]
\begin{center}
\parbox[t][][c]{0.2\textwidth}{\includegraphics[width=0.2\textwidth]{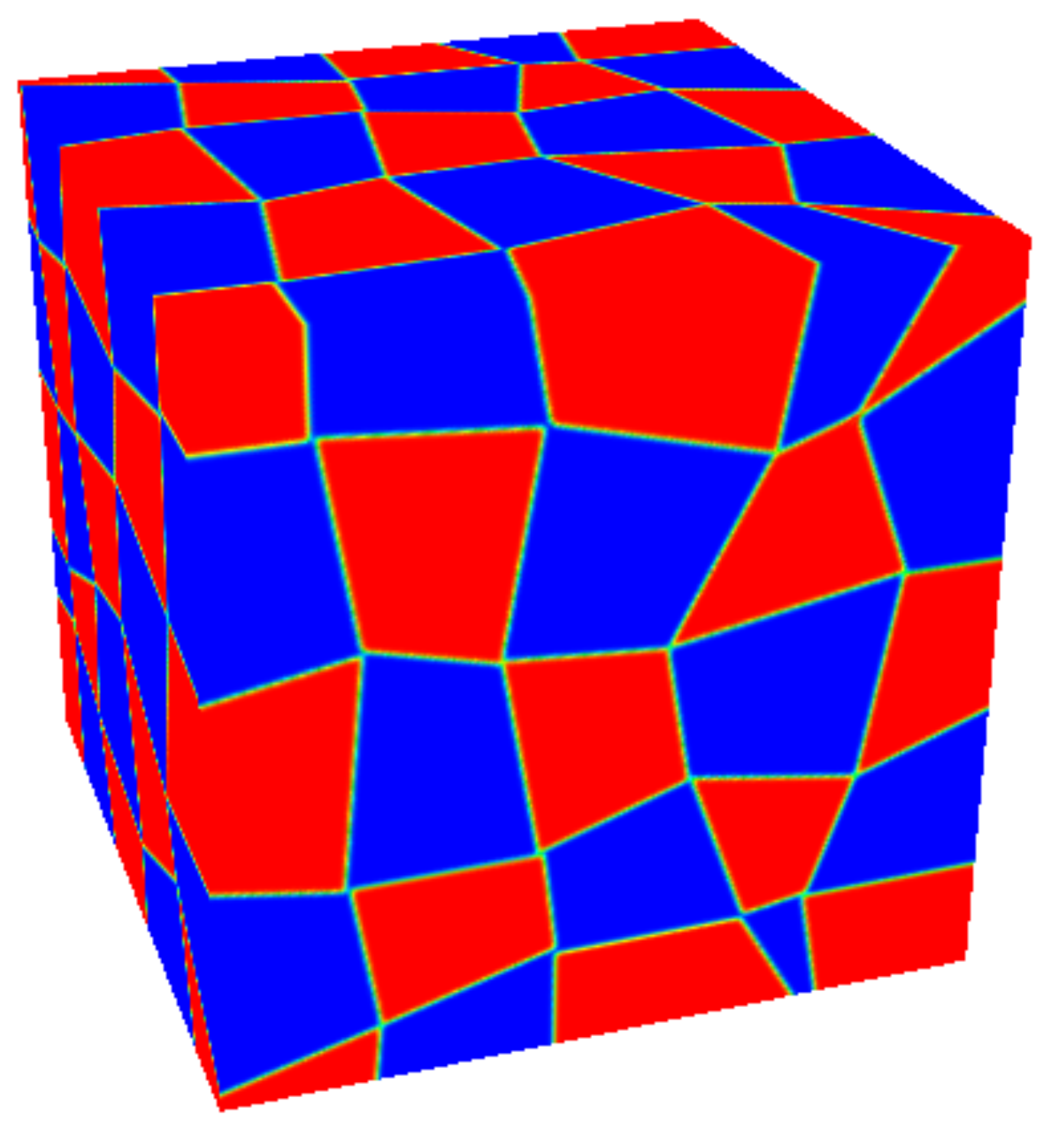}\\
 $\langle F^2\rangle=B^2$\\
$\langle| F\widetilde{F}|\rangle=B^2$}
\parbox[t][][c]{0.2\textwidth}{\includegraphics[width=0.2\textwidth]{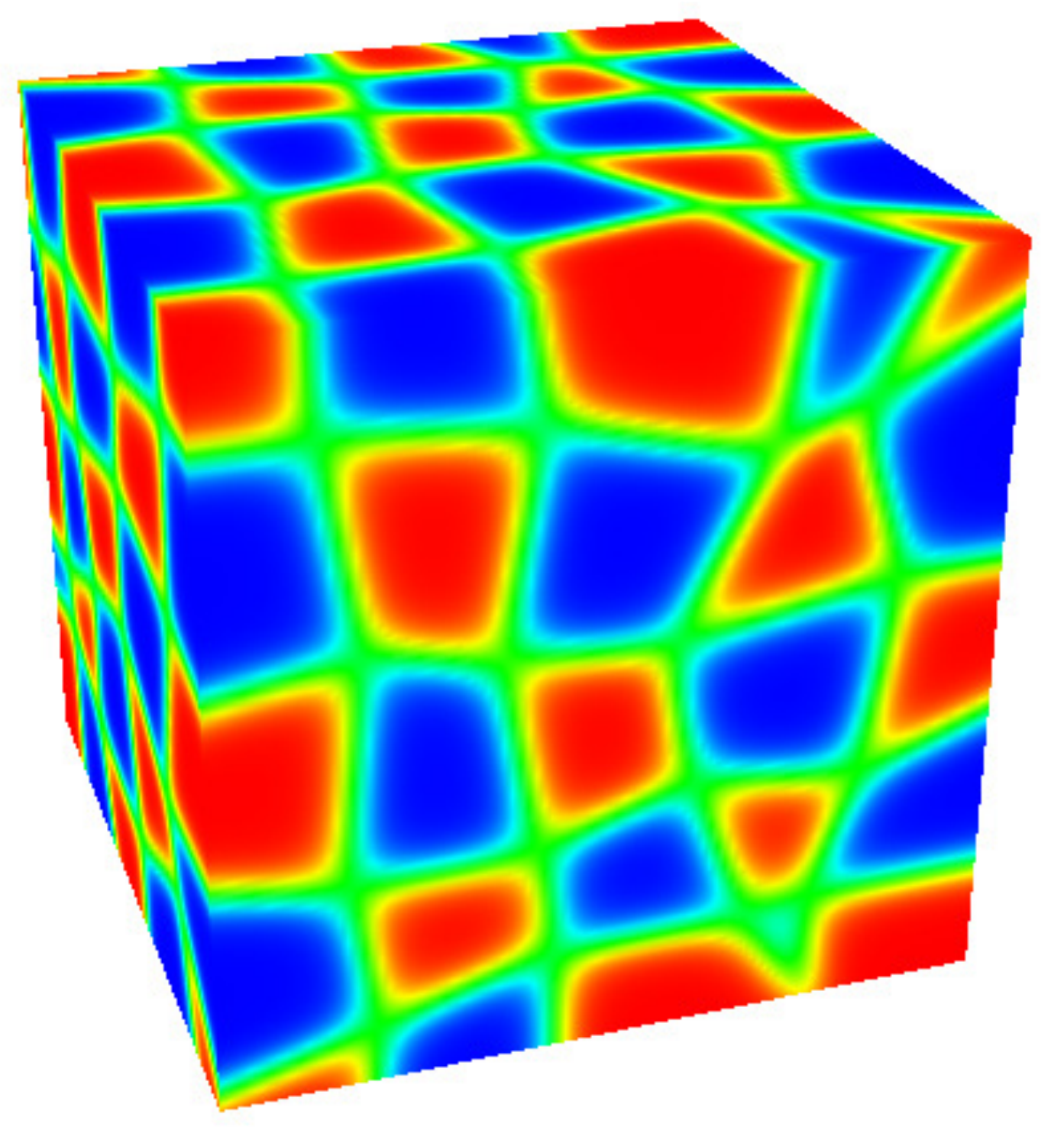}}
\parbox[t][][c]{0.2\textwidth}{\includegraphics[width=0.2\textwidth]{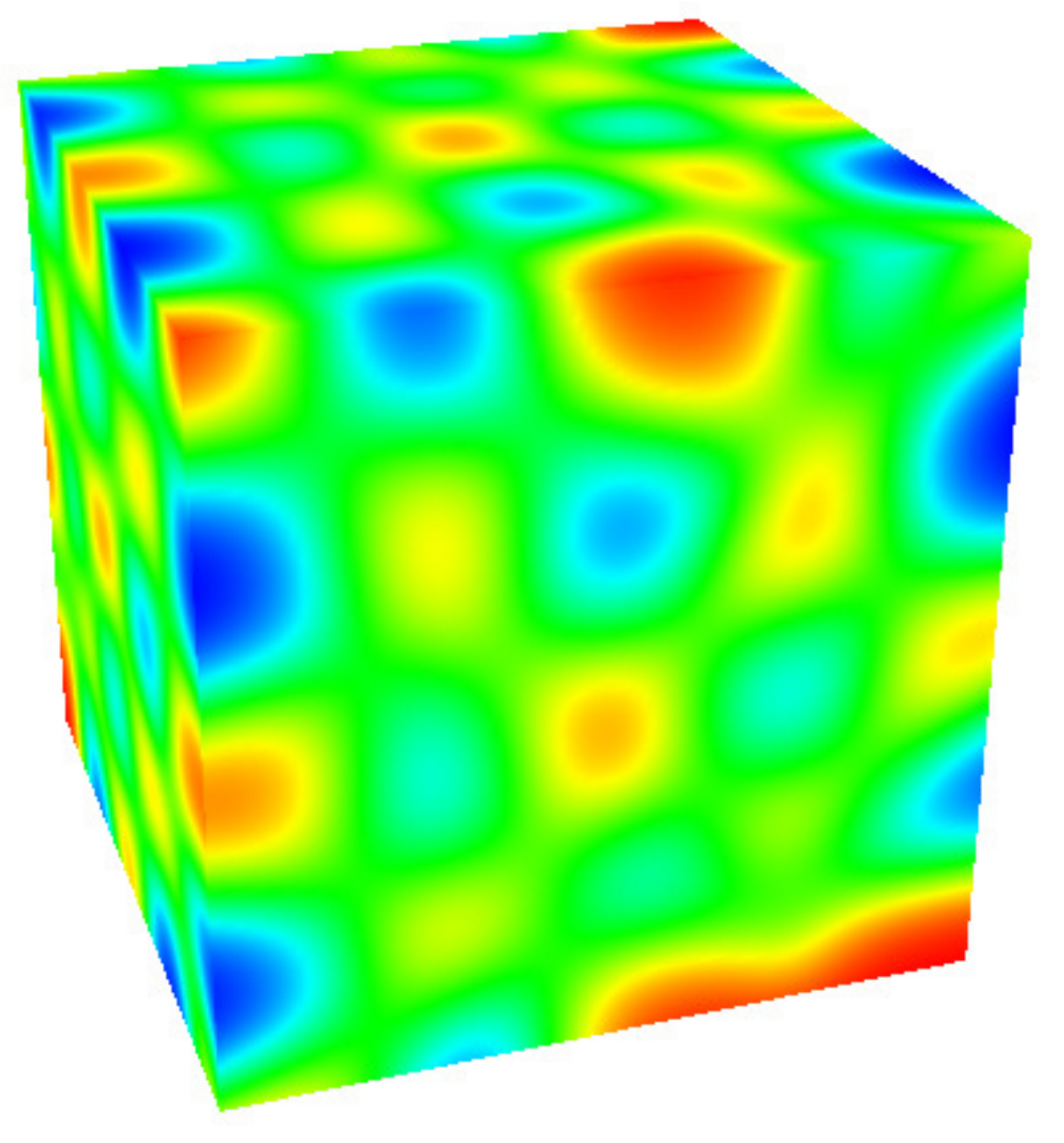}\\
$\langle F^2\rangle=B^2$\\
$\langle |F\widetilde{F}|\rangle\ll B^2$}
\end{center}
\caption{Topological charge density for domain wall networks with different values of $\mu$. The leftmost picture is an example of confining almost everywhere homogeneous Abelian (anti-)self-dual fields. Blue (red) color corresponds to the self-dual field (anti-self-dual), green - pure chromomagnetic field.\label{cubes}}
\end{figure}

Away from the kink location vacuum  field is almost self-dual ($\omega=0$) or anti-self-dual ($\omega=\pi$).  Exactly at the wall it becomes purely chromomagnetic ($\omega=0$).  Domain wall network is constructed  by means of the kink superposition.
In general kink can be parametrized as 
\begin{equation}
 \zeta(\mu_i,\eta_\nu^{i}x_\nu-q^{i})=\frac{2}{\pi}\arctan\exp(\mu_i(\eta_\nu^{i}x_\nu-q^{i})),
\end{equation}
where $\mu^i$ is inverse width of the kink, $\eta_\nu^i$ is a normal to the wall and $q^i=\eta_\nu^i x_\nu$ are coordinates of the wall.
A single lump in two, three and four dimensions is given by
\begin{equation}
 \label{asd_lump}
\omega(x)=\pi\prod_{i=1}^k \zeta(\mu_i,\eta_\nu^{i}x_\nu-q^{i}).
\end{equation}
for $k=4,6,8$, respectively.
The general kink network is then given by the additive superposition of lumps
\begin{equation}
 \label{kinknet}
\omega=\pi\sum_{j=1}^{\infty}\prod_{i=1}^k \zeta(\mu_{ij},\eta_\nu^{ij}x_\nu-q^{ij}).
\end{equation}
Topological charge density distribution for a network of domain walls with different width is illustrated in Fig.\ref{cubes}.

\section{Hadronization within the domain model of QCD vacuum \label{section_domain_model}}

The haronization formalism based on domain model of QCD vacuum  was elaborated in the series of papers \refcite{EN1,EN,NK1,NK4}. We refer to these papers for most of the technical details omitted in this brief presentation.  It has been shown that the  model  embraces static and dynamical quark confinement as well as spontaneous breaking of chiral symmetry. Estimation of masses of light, heavy-light mesons and heavy quarkonia along with their orbital excitations\cite{EN1,EN,NK4} demonstrated impressive phenomenological performance.  Calculations in Refs.~\refcite{EN,NK4} have been done neglecting certain mixing between  radially excited meson fields.  Below we present results of  calculation  refined in this respect.  

Nonperturbative gluon  fields are represented  in the model by the ensemble of domain-structured fields with the field strength tensor\cite{NK1,NK4}
\begin{gather}
\nonumber
F^{a}_{\mu\nu}(x)
=\sum_{k=1}^N n^{(k)a}B^{(k)}_{\mu\nu}\theta(1-(x-z_k)^2/R^2), \quad B^{(k)}_{\mu\nu}B^{(k)}_{\mu\rho}=B^2\delta_{\nu\rho}, \quad B=\frac{2}{\sqrt{3}}\Lambda^2,
\\\nonumber
\tilde B^{(k)}_{\mu\nu}=\pm B^{(k)}_{\mu\nu}, \quad 
\hat n^{(k)}=t^3\cos\xi_k+t^8\sin\xi_k, \quad \xi_k\in\left\{\frac{\pi}{6}(2l+1),\ l=0,\dots,5\right\},
\end{gather}
where scale $\Lambda$ and  mean domain radius $R$  are parameters of the model related to the
scalar gluon condensate and topological susceptibility  respectively\cite{NK1}.

\begin{table}[ph]
\tbl{Model parameters fitted to the masses of $\pi,\rho,K,K^*, \eta', J/\psi,\Upsilon$ and used in  calculation of all other quantities.}
{\begin{tabular}{@{}ccccccc@{}}
 \toprule
$m_{u/d}$, MeV&$m_s$, MeV&$m_c$, MeV&$m_b$, MeV&$\Lambda$, MeV&$\alpha_s$&$R$, fm\\
\colrule
$145$&$376$&$1566$&$4879$&$416$&$3.45$&$1.12$\\
\botrule
\end{tabular}}
\label{values_of_parameters}
\end{table}

The measure of  integration over ensemble of background fields is defined as
\begin{eqnarray*}
\label{measure}
\int\limits_{\cal B}dB\dots  &=& \prod_i
\frac{1}{24\pi^2}
\int_V\frac{d^4z_i}{V}
\int\limits_0^{2\pi}d\varphi_i\int_0^\pi d\theta_i\sin\theta_i
\\
&\times&\int_0^{2\pi} d\xi_i\sum\limits_{l=0,1,2}^{3,4,5}
\delta(\xi_i-\frac{(2l+1)\pi}{6})
\int_0^\pi d\omega_i\sum\limits_{k=0,1}\delta(\omega_i-\pi k)
\dots 
\end{eqnarray*}

Once measure of integration for background field is specified, one may return to functional integral of QCD  and, recalling the definition of Green's functions,
\begin{equation*}
G^{a_1\dots a_n}_{\mu_1\dots \mu_n}(x_1,\dots,x_n|B)=\left. \frac{1}{g^n}\frac{\delta^n \ln W[j]}{\delta j^{a_1}_{\mu_1}(x_1)\dots \delta j^{a_n}_{\mu_n}(x_n)} \right\lvert_{j=0},
\end{equation*}
 arrive at the representation 
\begin{eqnarray*}
{\cal Z}  &=&  
\int dB \int_{\Psi}{\cal D}\psi {\cal D}\bar \psi
\exp\left\{\int dx \bar\psi\left(i \hspace*{-0.2em} \not{\hspace*{-0.2em}\partial} + g\hspace*{-0.3em}\not{\hspace*{-0.3em}B}-m\right)\psi\right\} W[J],
\\
W[J]&=&\exp\left\{\sum_n \frac{g^n}{n!}\int d^4x_1\dots \int d^4x_n j^{a_1}_{\mu_1}(x_1)\dots j^{a_n}_{\mu_n}(x_n) G^{a_1\dots a_n}_{\mu_1\dots \mu_n}(x_1,\dots,x_n|B)\right\},
\end{eqnarray*}
where $j^a_\mu(x) =\bar\psi(x) \gamma_\mu t^a \psi(x)$ is the quark current.   By construction the gauge coupling constant $g$ here appears to be renormalized within some appropriate renormalization scheme, and the $n$-point correlators  are  exact renormalized  $n$-point gluon Green functions of pure gauge theory in the presence of the background field $B$.  Needless to say that exact form of the Green functions is not known, and certain reliable approximations have to be introduced. We truncate the exponent in $W[j]$ up to the four-fermion interaction term and approximate the two-point correlation function   by the gluon propagator in the presence of homogeneous Abelian (anti-)self-dual field neglecting radiative perturbative corrections (for details see Refs.\refcite{EN1,EN,NK4}). 
The randomness  of domain ensemble is taken into account implicitly by means of  averaging the quark loops over all possible configurations of the homogeneous background field at the final stage of derivation of the effective meson action\cite{EN,NK4}.

\begin{table}[ph]
\tbl{Masses of light mesons. $\tilde M$ denotes the value in the chiral limit.}
{\begin{tabular}{@{}ccccc|ccccc@{}} \toprule
Meson  & $n$ & $M_{\rm exp}$\cite{PDG} & $M$ & \rule{0ex}{1.2em}$\widetilde{M}$ &
 Meson & $n$ &  $M_{\rm exp}$\cite{PDG} & $M$ & \rule{0ex}{1.2em}$\widetilde{M}$\\
             &       &( MeV)                                   & (MeV)& (MeV)                                              & 
             &        &( MeV)                                 & (MeV)& (MeV)\\\colrule
$\pi$       &0 &140  &140  &0&$\rho$      &0 &775  &775  &769\\
$\pi(1300)$ &1 &1300 &1310 &1301&$\rho(1450)$&1 &1450 &1571 &1576\\
$\pi(1800)$ &1 &1812 &1503 &1466     &$\rho$      &2 &1720 &1946 &2098\\
\colrule
$K$              &0 &  494  &494  &      0      &$K^*$            &0 & 892  &892   &769\\
$K(1460)$   &1 &1460 &1302 &1301      &$K^*(1410)$ &1 &1410 &1443 &1576\\
$K$              &2 &         &1655 &1466      &$K^*(1717)$ &1 &1717 &1781 &2098\\
\colrule
$\eta$            &0& 548 & 621    &0        &$\omega$      &0       &775    &775   &769\\
$\eta'$           &0& 958 & 958     &872   &$\phi$            &0       &1019  &1039 &769\\
$\eta(1295)$&1& 1294 & 1138 &1361 &$\phi(1680)$ &1       &1680  &1686 &1576\\
$\eta(1475)$&1& 1476 & 1297 &1516 &$\phi$            &2       & 2175 &1897 &2098\\
\botrule
\end{tabular}}
\label{light_mesons}
\end{table}

Approximated functional integral reads
\begin{eqnarray}
\mathcal{Z}=\int_{\cal B} dB \int_{\Psi}{\cal D}\psi {\cal D}\bar \psi
\exp\left\{\int d^4x \bar\psi\left(i \hspace*{-0.2em} \not{\hspace*{-0.2em}\partial} + g\hspace*{-0.3em}\not{\hspace*{-0.3em}B}-m\right)\psi+\mathcal{L}\right\},
\label{4ferm}\\
\mathcal{L}=\frac{g^2}{2}\int d^4x \int d^4y\  G^{a b}_{\mu \nu}(x,y|B)j^{a}_{\mu}(x) j^{b}_{\nu}(y),
\nonumber
\end{eqnarray}
where $m$ is a diagonal quark mass matrix.

 By means of the Fiertz transformation of color, Dirac and flavour matrices the four-quark interaction can be rewritten as
 \begin{eqnarray}
\mathcal{L}=\frac{g^2}{2}\sum_{c,J}C_J\int d^4x \int d^4y  D(x-y)J^{Jc}(x,y|B)J^{Jc}(y,x|B) ,
\nonumber
\end{eqnarray}
 where numerical coefficients $C_J$ are different for different spin-parity  $J=S,P,V,A$.
Here bilocal color neutral quark currents
 \begin{equation*}
J^{Jc}(x,y|B)= \bar\psi^{if}_\alpha(x)\lambda_{ff'}^c\Gamma_J^{\alpha \alpha'}\left[\left\{\frac{ i}{2}x_\mu \hat B_{\mu\nu}y_\nu \right\}\right]^{ii'}\psi^{i'f'}_{\alpha'}(y)
\end{equation*}
are singlets with respect to the local background gauge transformations,  $\hat B_{\mu\nu}$ 
denotes background field strength in the fundamental representation, and
$i$, $f$ and $\alpha$ are color, flavour and space-time indices respectively. In the center of  quark mass coordinate system bilocal currents take the form
 \begin{eqnarray*}
&&J^{Jc}(x,y|B)\to J^{Jc}(x,z|B)=\bar\psi^{if}_\alpha(x)t_{ff'}^c\Gamma_J^{\alpha \alpha'}
\left[\exp\left\{ iz_\mu  \stackrel{\leftrightarrow}{\mathcal{D}}_\mu(x) \right\}\right]^{ii'}_{ff'}\psi^{i'f'}_{\alpha'}(x),
\\
&&\stackrel{\leftrightarrow}{\mathcal{D}}^{ff'}_{\mu}(x)=\xi_f\stackrel{\leftarrow}{\mathcal{D}}_{\mu}(x)-\xi_{f'}\stackrel{\rightarrow}{\mathcal{D}}_{\mu}(x), \ \ \xi_f=\frac{m_f}{m_f+m_{f'}},\ \xi_{f'}=\frac{m_f}{m_f+m_{f'}},
\end{eqnarray*}
and  their interaction is described by the action\cite{EN1}
 \begin{eqnarray}
&&\mathcal{L}=\frac{g^2}{2}\sum_{c,J}C_J\int d^4x \int d^4z D(z)J_{Jc}^{\dagger}(x,z|B)J_{Jc}(x,z|B),
\label{4ferm-1}\\
&& D(z)=\frac{1}{4\pi^2 z^2}\exp\left\{-\frac{1}{4}\Lambda^2z^2\right\}, 
\label{glprop-1}
\end{eqnarray}

\begin{table}[ph]
\tbl{Masses of heavy-light mesons and their lowest radial excitations .}
{\begin{tabular}{@{}cccc|cccc@{}} \toprule
Meson & $n$ & $M_{\rm exp}$\cite{PDG} & $M$ & Meson & $n$ & $M_{\rm exp}$\cite{PDG} & $M$\\
&&(MeV)& (MeV)&&&(MeV)& (MeV)\\
\colrule
$D$  &0 &1864      &1715 & $D^*$  &0 &2010        &1944\\
$D$  &1 &&2274& $D^*$  &1 &&2341\\
$D$  &2 &                        &2508  & $D^*$  &2 &                        &2564 \\\colrule
$D_s$&0 &1968       &1827 &  $D_s^*$&0 &2112       &2092 \\
$D_s$&1 &&2521 &  $D_s^*$&1 &&2578 \\
$D_s$&2 &&2808 & $D_s^*$&2 &                        &2859  \\\colrule
$B$  &0 &5279         &5041 &  $B^*$  &0 &5325         &5215 \\
$B$  &1 &&5535 &     $B^*$  &1 &&5578  \\
$B$  &2 &                        &5746 &  $B^*$  &2 &                        &5781 \\\colrule
$B_s$&0 &5366         &5135 &  $B_s^*$&0 &5415          &5355 \\
$B_s$&1 &&5746 & $B_s^*$&1 &&5783  \\
$B_s$&2 &                        &5988 & $B_s^*$&2 &                        &6021  \\\colrule
$B_c$&0 &6277          &5952 &  $B_c^*$&0 &                        &6310 \\
$B_c$&1 &&6904 &  $B_c^*$&1 &                        &6938 \\
$B_c$&2 &                        &7233 &  $B_c^*$&2 &                        &7260 \\
 \botrule
\end{tabular} }
\label{heavy-light}
\end{table}

\noindent
where $x_\mu$ - center of mass coordinates, and $z_\mu$ - relative coordinates of quark and antiquark
and  $\Lambda^2=\sqrt{3}B/2$.
Function   $D(z)$ originates from the gluon propagator in the presence of the homogeneous Abelian (anti-)self-dual gluon field. It differs from the free massless scalar propagator by the Gaussian exponent,
which completely changes the IR properties of the propagator. In momentum representation it takes the form
\begin{eqnarray*}
G\left(p^2\right) = \frac{1}{p^2}\left(1-e^{-p^2/\Lambda^2}\right).
\end{eqnarray*}

It is important that the gluon condensate encoded in the background vacuum field under consideration removes the pole from the propagator which can be treated as dynamical confinement. The same property is characteristic for the quark propagator in the homogeneous as well as domain structured\cite{NK1} Abelian (anti-)self-dual gluon field. 
Momentum representation $\widetilde H_f(p|B)$ of the translation invariant part of the quark propagator in the presence of the homogeneous background field, 
\begin{equation*}
S(x,y)=\exp\left(-\frac{i}{2}x_\mu B_{\mu\nu}y_\nu\right)H(x-y),
\end{equation*}
is an entire analytical function of momentum:
\begin{multline*}
\widetilde H_f(p)=\frac{1}{2\upsilon \Lambda^2} \int_0^1 ds e^{(-p^2/2\upsilon \Lambda^2)s}\left(\frac{1-s}{1+s}\right)^{m_f^2/4\upsilon \Lambda^2}\left[\vphantom{\frac{s}{1-s^2}}p_\alpha\gamma_\alpha\pm is\gamma_5\gamma_\alpha f_{\alpha\beta} p_\beta+\right.\\
\left.\mbox{}+m_f\left(P_\pm+P_\mp\frac{1+s^2}{1-s^2}-\frac{i}{2}\gamma_\alpha f_{\alpha\beta}\gamma_\beta\frac{s}{1-s^2}\right)\right],
\end{multline*}
\begin{gather*}
f_{\alpha\beta}=\frac{\hat{n}}{2\upsilon\Lambda^2}B_{\alpha\beta}, \upsilon=\mathrm{diag}\left(\frac16,\frac16,\frac13\right).
\end{gather*}

\begin{table}[ph]
\tbl{Masses of heavy quarkonia.}
{\begin{tabular}{@{}cccc@{}} \toprule
Meson&$n$&$M_{\rm exp}$\cite{PDG}&$M$\\
&&(MeV) & (MeV)\\
\colrule
$\eta_c(1S)$  &0 &2981 &2751\\
$\eta_c(2S)$  &1 &3639 &3620\\
$\eta_c$      &2 &     &3882\\
\colrule
$J/\psi(1S)$  &0 &3097 &3097\\
$\psi(2S)$    &1 &3686 &3665\\
$\psi(3770)$  &2 &3773 &3810\\
\colrule
$\Upsilon(1S)$&0 &9460 &9460\\
$\Upsilon(2S)$&1 &10023&10102\\
$\Upsilon(3S)$&2 &10355&10249\\\botrule
\end{tabular} \label{ta1}}
\label{heavy}
\end{table}

There are two equivalent ways to derive effective meson action based on the functional integral  \eqref{4ferm} with the interaction term $\mathcal{L}$  taken in the form~\eqref{4ferm-1}.  The first one is a bosonization of the functional integral in terms of bilocal meson-like fields (see for example Ref.~\refcite{roberts}). We shall return to this option in the discussion section. Another, more transparent in our opinion way is to decompose the bilocal currents 
over complete set of  functions $f^{nl}_{\mu_1\dots \mu_l}(z)$ characterized by the radial quantum number $n$ and orbital momentum $l$  and orthogonal with  the weight determined by the gluon propagator \eqref{glprop-1} in Eq.\eqref{4ferm-1},
\begin{eqnarray}
J^{aJ}(x,z)=\sum_{nl}\left(z^2\right)^{l/2} f^{nl}_{\mu_1\dots \mu_l}(z) J^{aJln}_{\mu_1\dots \mu_l}(x).
\nonumber
\end{eqnarray}

The choice of functions  $f^{nl}(z)$ is unambiguously determined by the gluon propagator
\begin{eqnarray}
f^{nl}_{\mu_1\dots \mu_l}=L_{nl}\left(z^2\right) T^{(l)}_{\mu_1\dots \mu_l}(n_z),\quad n_z=z/\sqrt{z^2}, 
\label{wavef}
\end{eqnarray}
where $T^{(l)}_{\mu_1\dots \mu_l}$ are irreducible tensors of four-dimensional rotational group, and 
generalized Laguerre polynomials $L_{nl}$  obey relation
\begin{equation}
\int_0^\infty du \rho_l(u)L_{nl}(u)L_{n'l}(u)=\delta_{nn'},\quad \rho_l(u)=u^{l}e^{-u}.
\end{equation}

\begin{table}[ph]
\tbl{Decay and transition constants of various  mesons}
{\begin{tabular}{@{}cccc|cccc@{}} \toprule
Meson&$n$&$f_P^{\rm exp}$&$f_P$&Meson&$n$&$g_{V\gamma}$ \cite{PDG}&$g_{V\gamma}$\\
&&(MeV)& (MeV)&&&\\
\colrule
$\pi$      &0 &130 \cite{PDG}        &140 & $\rho$&0&0.2&0.2 \\
$\pi(1300)$&1 &---                   &29 & $\rho$&1&&0.034 \\
\colrule
$K$        &0 &156 \cite{PDG}        &175  & $\omega$&0&0.059&0.067\\
$K(1460)$  &1 &---                   &27   & $\omega$&1&&0.011\\
\colrule
$D$        &0 &205 \cite{PDG}        &212  & $\phi$&0&0.074&0.069\\
$D$        &1 &---                   &51   & $\phi$&1&&0.025\\
\colrule
$D_s$      &0 &258 \cite{PDG}        &274  & $J/\psi$&0&0.09&0.057\\
$D_s$      &1 &---                   &57  &  $J/\psi$&1&&0.024\\
\colrule
$B$        &0 &191 \cite{PDG}        &187  & $\Upsilon$&0&0.025&0.011\\
$B$        &1 &---                   &55   & $\Upsilon$&1&&0.0039\\
\colrule
$B_s$      &0 &253 \cite{Chiu:2007bc}&248  &  & &\\
$B_s$      &1 &---                   &68   &  & &\\
\colrule
$B_c$      &0 &489 \cite{Chiu:2007bc}&434  &  & &\\
$B_c$      &1 &                  &135  & &&\\
\botrule
\end{tabular}
\label{constants}}
\end{table}

\noindent
The weight $\rho_l(u)$ originates from the gluon propagator \eqref{glprop-1}.  Nonlocal quark currents $J^{aJln}_{\mu_1\dots \mu_l}$ with  complete set of meson quantum numbers depend only on the center of mass coordinate $x$\cite{EN,NK4},
\begin{eqnarray}
J^{aJln}_{\mu_1\dots \mu_l}(x)=\bar{q}(x)V^{aJln}_{\mu_1\dots \mu_l}\left(\frac{\stackrel{\leftrightarrow}{D} \!
(x)}{\Lambda}\right)q(x),
\nonumber\\
V^{aJln}_{\mu_1\dots\mu_l}(x)=M^a\Gamma^J F_{nl}\left(\frac{\stackrel{\leftrightarrow}{D}^2\!\!\!
(x)}{\Lambda^2}\right)T^{(l)}_{\mu_1\dots\mu_l}\left(\frac{1}{i}\frac{\stackrel{\leftrightarrow}{D}\!(x)}{\Lambda}\right),
\label{qmvert}\\
F_{nl}(s)=s^n\int_0^1 dt t^{n+l} \exp(st),
\nonumber
\end{eqnarray}
and the four-fermion interaction takes the form
\begin{eqnarray*}
\mathcal{L}=\frac{g^2}{2} \sum_{a,J,l,n} C_J \int d^4x J_{aJln}^{\dagger}(x)J_{aJln}(x).
\end{eqnarray*}

Further technical details of decomposition of bilocal quark currents and bosonization can be found in Ref.~\refcite{EN,NK4}. By means of  bosonization the truncated QCD functional integral can be rewritten in terms of the   composite colorless meson fields:
\begin{gather}
\label{effective_meson_action}
Z={\cal N}
\int D\Phi_{\cal Q}
\exp\left\{-\frac{\Lambda}{2}\frac{h^2_{\cal Q}}{g^2 C_{\cal Q}}\int dx 
\Phi^2_{\cal Q}(x)
-\sum\limits_k\frac{1}{k}W_k[\Phi]\right\},
\end{gather}
where $\mathcal{Q}$ denotes all meson quantum numbers.
 Meson-meson interactions are  described by by $k$-point nonlocal vertices $\Gamma^{(k)}_{\mathcal{Q}_1\dots \mathcal{Q}_2}$:
\begin{gather*}
W_k[\Phi]=
\sum\limits_{{\cal Q}_1\dots{\cal Q}_k}h_{{\cal Q}_1}\dots h_{{\cal Q}_k}
\int dx_1\dots\int dx_k
\Phi_{{\cal Q}_1}(x_1)\dots \Phi_{{\cal Q}_k}(x_k)
\Gamma^{(k)}_{{\cal Q}_1\dots{\cal Q}_k}(x_1,\dots,x_k)\\
\Gamma^{(2)}_{{\cal Q}_1{\cal Q}_2}=
\overline{G^{(2)}_{{\cal Q}_1{\cal Q}_2}(x_1,x_2)}-
\Xi_2(x_1-x_2)\overline{G^{(1)}_{{\cal Q}_1}G^{(1)}_{{\cal Q}_2}},
\\\nonumber
\Gamma^{(3)}_{{\cal Q}_1{\cal Q}_2{\cal Q}_3}=
\overline{G^{(3)}_{{\cal Q}_1{\cal Q}_2{\cal Q}_3}(x_1,x_2,x_3)}-
\frac{3}{2}\Xi_2(x_1-x_3)
\overline{G^{(2)}_{{\cal Q}_1{\cal Q}_2}(x_1,x_2)
G^{(1)}_{{\cal Q}_3}(x_3)}
\\
+
\frac{1}{2}\Xi_3(x_1,x_2,x_3)
\overline{G^{(1)}_{{\cal Q}_1}(x_1)G^{(1)}_{{\cal Q}_2}(x_2)
G^{(1)}_{{\cal Q}_3}(x_3)}.
\end{gather*}
Vertices are expressed via quark loops $G^{(k)}_{{\cal Q}_1\dots{\cal Q}_k}$:
\begin{equation*}
\overline{G^{(k)}_{{\cal Q}_1\dots{\cal Q}_k}(x_1,\dots,x_k)}
=\int d B_j
{\rm Tr}V_{{\cal Q}_1}\left(x_1|B^{(j)}\right)S\left(x_1,x_2\right)
V_{{\cal Q}_k}\left(x_k|B^{(j)}\right)S\left(x_k,x_1\right),
\end{equation*}
\begin{multline*}
\overline{G^{(l)}_{{\cal Q}_1\dots{\cal Q}_l}(x_1,\dots,x_l)
G^{(k)}_{{\cal Q}_{l+1}\dots{\cal Q}_k}(x_{l+1},\dots,x_k)}
=
\\
\int d B_j
{\rm Tr}\left\{
V_{{\cal Q}_1}\left(x_1|B^{(j)}\right)S\left(x_1,x_2|B^{(j)}\right)\dots
V_{{\cal Q}_k}\left(x_l|B^{(j)}\right)S\left(x_l,x_1|B^{(j)}\right)
\right\}\\
\times
{\rm Tr}\left\{
V_{{\cal Q}_{l+1}}\left(x_{l+1}|B^{(j)}\right)S\left(x_{l+1},x_{l+2}|B^{(j)}\right)\dots
V_{{\cal Q}_k}\left(x_k|B^{(j)}\right)S\left(x_k,x_{l+1}|B^{(j)}\right)
\right\}.
\end{multline*}

Bar denotes integration over all configurations of the background fields.   Meson vertex functions include in general  several  one-loop diagrams  correlated via the background field.  Geometrically, the $n$-point correlator $\Xi_n(x_1,\dots,x_n)$ is given by a volume of overlap of $n$ four-dimensional hyperspheres\cite{NK4}.

The mass spectrum $M_\mathcal{Q}$  of mesons  and quark-meson coupling constants  $h_{\cal Q}$ are determined by the quadratic part of the effective meson action via equations
\begin{gather}
\label{mass-eq}
1=
\frac{g^2C_{\cal Q}}{\Lambda^2}\tilde \Pi_{\cal Q}(-M^2_{\cal Q}|B),
\ \ \
h^{-2}_{\cal Q}=
\frac{d}{dp^2}\tilde\Pi_{\cal Q}(p^2)|_{p^2=-M^2_{\cal Q}}, 
\end{gather}
where $\tilde\Pi_{\cal Q}(p^2)$ are the eigenvalues of the two-point vertex $\tilde\Gamma^{(2)}_{\cal QQ'}(p)$ .

The free parameters of the model are the strong coupling constant $\alpha_s$, quark masses  $m_{u/d}$, $m_s$, $m_c$, $m_b$, the scale $\Lambda$ and mean domain size $ R$.
By construction the strong coupling constant and the quark masses  correspond to the infrared limit of the running  renormalized quantities.  The scale  and mean domain size are related to the scalar gluon condensate and topological susceptibility of  pure gluodynamics.

\section{Results for decay constants and masses of mesons \label{section_masses_of_mesons}}
Dependence  of the masses  on radial quantum number $n$ and orbital momentum $l$  has Regge character as it has been shown long time ago\cite{EN1}.
The source of Regge mass spectrum in the model is the same as the source of dynamical color confinement -- Gaussian exponents in the gluon and quark propagators due to the impact of the Abelian (anti-)self-dual background fields. 
In our previous estimations\cite{EN,NK4}, nondiagonal in radial number terms in the quadratic part of the action were neglected. Results of  improved computation with proper diagonalisation is given in Tables~\ref{light_mesons}, \ref{heavy-light} and \ref{heavy}. 
The values of parameters given in Table \ref{values_of_parameters} were fitted to ground state mesons $\pi,\rho,K, K^*,\eta', J/\psi,\Upsilon$. The rest of masses were computed straightforwardly without further tuning of the parameters. In particular the same strong coupling constant was taken for light, heavy light mesons and heavy quarkonia. Computed values of  leptonic decay constants $f_P$ and $V\to \gamma$ transition constants $g_{V\gamma}$ are given in Table~\ref{constants}. In comparison with available experimental data the overall accuracy of the model is about 11-15\% (with very few exceptions like  $g_{V\gamma}$ for heavy quarkonia).

\section{Discussion}

Bosonization of the four-quark interaction \eqref{4ferm-1} in terms of bilocal  meson-like fields
$\Phi_{Jc}(x,z)$ leads to the following quadratic part of the effective action
\begin{eqnarray}
&&\mathcal{S}_2=-\frac{1}{2}\int d^4x\int d^4z D(z)\Phi^2_{Jc}(x,z) 
\nonumber\\
&&-
2g^2\int d^4x d^4x' d^4z d^4z'  D(z)D(z')  
\Phi_{Jc}(x,z)\Pi_{Jc,J'c'}(x,x';z,z')\Phi_{J'c'}(x',z'),
\nonumber\\ 
&&\Pi_{Jc,J'c'}(x,x';z,z')={\rm Tr} V_{Jc}(x,z) S(x,x') V_{J'c'}(x',z') S(x',x),
\nonumber\\
&& V_{Jc}(x,z)=\Gamma_J t_c\exp\left\{ iz_\mu  \stackrel{\leftrightarrow}{\mathcal{D}}_\mu(x) \right\}.
\label{bilocm}
\end{eqnarray}
 Meson wave functions $f^{nl}_{\mu_1\dots \mu_l}(z)$ in the decomposition 
$$
\Phi^{aJ}(x,z)=\sum_{nl}\left(z^2\right)^{l/2} f^{nl}_{\mu_1\dots \mu_l}(z) \Phi^{aJln}_{\mu_1\dots \mu_l}(x)
$$
are defined by the action $S_2$  \textit{via } corresponding integral equation. Solution of this  eigenfunction  problem is equivalent to diagonalization of the quadratic part  of the effective action in Eq.~\eqref{effective_meson_action}. Specific Gaussian form \eqref{glprop-1} of gluon propagator $D(z)$ is the reason for the radial part of the wave functions to be represented by the generalized Laguerre polynomials, see Eq.~\eqref{wavef}. The mass spectrum has Regge character.
Effective action \eqref{bilocm} has to be compared with the soft wall AdS/QCD action.
For the quadratic in $z$ dilaton profile $\varphi(z)=\kappa z^2$ one arrives at the decomposition
\begin{eqnarray*}
\Phi_J(x,z)=\sum_n \phi_{nJ}(z)\Phi_{nJ}(x)
\end{eqnarray*}
with the radial meson wave functions proportional to generalized Laguerre polynomials,
$
\phi_{nJ}=R^{J-3/2}\kappa^{1+l}z^{l-J+2} L^l_n(\kappa^2z^2)
$,
which is a solution of the corresponding eigenvalue problem in  differential form. In this case the meson masses correspond strictly to Regge spectrum.

There are obvious differences between two actions.  There are four  $z$-coordinates in  \eqref{bilocm} and, hence, the meson wave function contains angular part, and it is not immediately clear where 
AdS metrics could come from in representation  \eqref{bilocm}. Apart from this, the feature in common is  the Gaussian  weight function in the actions which  in both cases plays the most important role for phenomenological output.  
One gets an impression that the form of dilaton profile in soft wall AdS/QCD approach can be linked to the gluon propagator and thus to the properties of QCD vacuum.
So far this observation looks as superficial one and requires more profound consideration.

\end{document}